\documentclass[prb,preprint,showpacs]{revtex4}

\usepackage{amsmath}
\usepackage{graphicx}
\usepackage{dcolumn}

\begin{document}

\title{Simple variational approaches to quantum wells}
\author{Francisco M. Fern\'{a}ndez}\email{fernande@quimica.unlp.edu.ar}

\affiliation{INIFTA (UNLP, CCT La Plata--CONICET), Blvd. 113 y 64 S/N, \\
Sucursal 4, Casilla de Correo 16, 1900 La Plata, Argentina}

\begin{abstract}
We discuss two simple variational approaches to quantum wells. The
trial harmonic functions analyzed in an earlier paper give
reasonable results for all well depths and are particularly
suitable for deep wells. On the other hand, the exponential
functions proposed here are preferable for shallow wells. We
compare the shallow-well expansions for both kind of functions and
show that they do not exhibit the cubic term appearing in the
exact series. It is also shown that the deep-well expansion for
the harmonic functions agree with the first terms of perturbation
theory.
\end{abstract}


\maketitle

\section{Introduction}

\label{sec:intro}

In a recent paper we discussed the application of the variational
method to a Gaussian well.\cite{F11a} We showed that harmonic
variational functions yield reasonably accurate results for all
well depths despite the fact that the shallow-well expansion for
the ground-state variational energy does not agree with the exact
expansion beyond the leading term.

Students at introductory courses on quantum mechanics are commonly
encouraged to solve the Schr\"odinger equation for a particle in a
one-dimensional square box with finite walls. They thus learn that
the exact solution decays exponentially as $|x|\rightarrow \infty
$. After such an experience they would find it unreasonable the
choice of harmonic variational functions that do not exhibit the
behavior at infinity expected for the exact solutions to the
Schr\"{o}dinger equation for the Gaussian well. For this reason we
think that it is interesting to show an alternative variational
calculation based on trial functions with the expected exponential
behavior at infinity.

In Sec.~\ref{sec:variational} we develop the variational method
for a general single-well potential-energy function. In Sec.~\ref
{sec:Gaussian_well} we discuss the application of the harmonic and
exponential trial functions to the Gaussian well, compare the
approximate and exact energies for the ground and first-excited
state as well as their shallow-well expansions. In
Sec.~\ref{sec:deep_well} we compare the variational and exact
deep-well expansions that we did not consider in our previous
paper.\cite{F11a} Finally, in Sec.~\ref{sec:conclusions} we
discuss the main results and draw conclusions.

\section{Simple variational method}

\label{sec:variational}

We are interested in the solutions to the Schr\"{o}dinger equation
\begin{equation}
\hat{H}\psi _{n}=\epsilon _{n}\psi _{n},\,n=0,1,\ldots  \label{eq:Schro}
\end{equation}
for the dimensionless Hamiltonian operator
\begin{equation}
\hat{H}=\hat{T}+v(x)=-\frac{1}{2}\frac{d^{2}}{dx^{2}}+v(x)
\label{eq:Hamiltonian}
\end{equation}
In our previous paper we showed how to convert the Schr\"{o}dinger
equation into a dimensionless eigenvalue equation and we do not
repeat it here.\cite {F11a} For concreteness we choose the
dimensionless potential-energy function to be of the form
$v(x)=-v_{0}f(x)$, where $v_{0}>0$ and the function $f(x)$
exhibits a maximum at $x=0$ and tends to zero when
$|x|\rightarrow\infty $. Since we restrict ourselves to single
wells we also assume that $f^{\prime }(x)>0$ if $x<0$ and
$f^{\prime }(x)<0$ if $x>0$ although we do not make use of this
condition explicitly. We will discuss a particular example later
on. The reader will find some references about quantum-mechanical
wells in our earlier paper.\cite{F11a}

It is our purpose to obtain simple solutions for some states of
the quantum well by means of the variational method. To this end
we choose a variational function $\varphi (a,x)$ that depends on a
variational parameter $a$. If we assume that $\epsilon
_{0}<\epsilon _{1}<\epsilon _{2}<\cdots $ then the variational
principle gives us an upper bound to the smallest eigenvalue
\begin{equation}
W(a)=\frac{\left\langle \varphi \right| \hat{H}\left| \varphi \right\rangle
}{\left\langle \varphi \right| \left. \varphi \right\rangle }\geq \epsilon
_{0}  \label{eq:var_princ}
\end{equation}
and we realize that the optimum value of $a$ is given by the minimum of $%
W(a) $. If we write $W(a)=F(a)-v_{0}G(a)$ where $F(a)=\left\langle \varphi
\right| \hat{T}\left| \varphi \right\rangle /\left\langle \varphi \right|
\left. \varphi \right\rangle $ and $G(a)=\left\langle \varphi \right|
f(x)\left| \varphi \right\rangle /\left\langle \varphi \right| \left.
\varphi \right\rangle $ then $W^{\prime }(a)=0$ becomes
\begin{equation}
F^{\prime }(a)-v_{0}G^{\prime }(a)=0  \label{eq:min_cond}
\end{equation}
Commonly, solving this equation for $a$ may not be possible or may lead to
cumbersome expressions. In that case we can resort to a simple parametric
expression for the approximate energy in the form
\begin{eqnarray}
v_{0} &=&\frac{F^{\prime }(a)}{G^{\prime }(a)}  \nonumber \\
W(a) &=&\frac{F(a)G^{\prime }(a)+F^{\prime }(a)G(a)}{G^{\prime }(a)}
\label{eq:W(a),v0(a)}
\end{eqnarray}

If $f(-x)=f(x)$ then the states have definite parity $\psi
_{n}(-x)=(-1)^{n}\psi _{n}(x)$ and the variational principle applies to $%
\epsilon _{0}$ or $\epsilon _{1}$ if the variational function is even or
odd, respectively. Therefore, the variational equations (\ref{eq:W(a),v0(a)}%
) apply to both the ground and first excited states.

\section{The Gaussian well}

\label{sec:Gaussian_well}

In order to illustrate the performance of the variational method on
quantum-mechanical wells we choose a particularly simple example that allows
us to calculate the functions $F(a)$ and $G(a)$ analytically. As in our
earlier paper we select the Gaussian well given by
\begin{equation}
f(x)=e^{-x^{2}}  \label{eq:f(x)_Gauss}
\end{equation}
and begin present discussion with the calculation based on
Harmonic-oscillator-like variational functions.\cite{F11a} The
simplest variational function for the ground state is
\begin{equation}
\varphi _{0h}(a,x)=e^{-ax^{2}}  \label{eq:varphi_0h}
\end{equation}
and we easily obtain

\begin{eqnarray}
v_{0} &=&\frac{\sqrt{2}\sqrt{a}\left( 2a+1\right) ^{3/2}}{2}  \nonumber \\
W_{0h}(a) &=&-\frac{a\left( 4a+1\right) }{2}  \label{eq:W0h(a)}
\end{eqnarray}

We showed that $W_{0h}\rightarrow 0$ as $v_{0}\rightarrow 0$ according to%
\cite{F11a}
\begin{equation}
W_{0h}=-v_{0}^{2}+4v_{0}^{4}-48v_{0}^{6}+832v_{0}^{8}-17408v_{0}^{10}+\ldots
\,.  \label{eq:W0h_weak}
\end{equation}
while the exact expansion is\cite{F11a}
\begin{equation}
\epsilon _{0}=-\frac{\pi v_{0}^{2}}{2}+\sqrt{2}\,\pi v_{0}^{3}-\frac{\pi
\left( 2\pi +3\sqrt{3}+3\right) v_{0}^{4}}{3}+\frac{\sqrt{2}\,\pi \left(
2\pi +3\sqrt{3}\right) v_{0}^{5}}{3}-\ldots \,.  \label{eq:e_weak}
\end{equation}
Note that the cubic term is missing in the approximate expansion (\ref
{eq:W0h_weak}).

We have also shown that
\begin{equation}
\varphi _{1h}(a,x)=xe^{-ax^{2}}  \label{eq:varphi_1h}
\end{equation}
is a suitable variational function for the first excited state,
and in this case we have

\begin{eqnarray}
v_{0} &=&\frac{\sqrt{2}\sqrt{2a+1}\left( 4a^{2}+4a+1\right) }{4\sqrt{a}}
\nonumber \\
W_{1h}(a) &=&-\frac{a\left( 8a^{2}+2a-1\right) }{2\left(
2a+1\right) } \label{eq:W1h(a)}
\end{eqnarray}

It is well known that $\epsilon _{1}(v_{0})\rightarrow 0^-$ as
$v_{0}$ approaches a critical well strength $v_{0,1}\approx1.342$
from above\cite {F11a,F11b}. We can estimate the critical strength
by means of equations (\ref{eq:W1h(a)}) in a quite simple way. We
first obtain a positive root of $W_{1h}(a_{c})=0$ and then
$v_{0}(a_{c})$; the result is $a_{c}=1/4$ and
$v_{0}(1/4)=9\sqrt{3}/8\approx 1.95$. Since the variational energy
is an upper bound to the exact one for all $v_{0}$ then it is not
surprising that $v_{0}(a_{c})>v_{0,1}$. The error for the critical
strength obtained with this variational function is rather large:
$45\%$.

We may try to improve those results by means of more convenient
trial functions. Since $v(x\rightarrow \pm \infty )=0$ we know
that the eigenfunctions behave as $\psi \sim e^{-\alpha |x|}$ for
sufficiently large $|x|$, where $\alpha =\sqrt{-2\epsilon }$. In
order to avoid the function $|x|$ in our calculations we work on
the half positive line $x>0$ and take into account the boundary
conditions at origin: $\psi (0)\neq 0$, $\psi ^{\prime }(0)=0$ for
the even states and $\psi (0)=0$, $\psi ^{\prime }(0)\neq 0$ for
the odd ones. Thus, the simplest trial exponential function for
the ground state appears to be
\begin{equation}
\varphi _{0e}(a,x)=(1+ax)e^{-ax},\,x>0  \label{eq:varphi_0e}
\end{equation}
The calculation of the integrals is straightforward and we obtain

\begin{eqnarray}
v_{0} &=&\frac{a}{\sqrt{\pi }e^{a^{2}}\left( 4a^{6}+4a^{4}-5a^{2}+2\right)
[1-\mathrm{erf}{\left( a\right) ]}-2a\left( 2a^{4}+a^{2}-2\right) }
\nonumber \\
W_{0e}(a) &=&\frac{a^{2}\left\{ \sqrt{\pi }e^{a^{2}}\left(
4a^{6}+a^{2}-2\right) \left[ \mathrm{erf}{\left( a\right) }-1\right]
+2a\left( 2a^{4}-a^{2}+2\right) \right\} }{10\left\{ \sqrt{\pi }%
e^{a^{2}}\left( 4a^{6}+4a^{4}-5a^{2}+2\right) \left[ \mathrm{erf}{\left(
a\right) }-1\right] +2a\left( 2a^{4}+a^{2}-2\right) \right\} }
\label{eq:W_1e}
\end{eqnarray}
where $\mathrm{erf}(z)$ is the error function.

In order to obtain the $v_{0}$-series for $W_{0e}$ we first expand $v_{0}(a)$
into an $a$-series:

\begin{equation}
v_{0}=\frac{a}{2\sqrt{\pi }}+\frac{3a^{3}}{4\sqrt{\pi }}-\frac{4a^{4}}{3\pi }%
+\frac{9a^{5}}{8\sqrt{\pi }}+\ldots  \label{eq:v0(a)_series_exp}
\end{equation}
that we invert to obtain the $v_{0}$-series for $a$

\begin{equation}
a=2\sqrt{\pi }v_{0}-12\pi ^{3/2}v_{0}^{3}+\frac{128\pi ^{3/2}}{3}%
v_{0}^{4}+144\pi ^{5/2}v_{0}^{5}+\ldots  \label{eq:a(v0)_series_exp}
\end{equation}
Then, we expand $W_{0e}$ in a Taylor series about $a=0$

\begin{equation}
W_{0e}=-\frac{a^{2}}{10}-\frac{a^{4}}{5}+\frac{2a^{5}}{5\sqrt{\pi }}+\ldots
\label{eq:W_0e_a-series_exp}
\end{equation}
and substitute the series (\ref{eq:a(v0)_series_exp}) to obtain the
shallow-well expansion

\begin{equation}
W_{0e}=-\frac{2\pi v_{0}^{2}}{5}+\frac{8\pi ^{2}v_{0}^{4}}{5}-\frac{64\pi
^{2}v_{0}^{5}}{15}+\ldots   \label{eq:W0e_weak}
\end{equation}
The calculation is straightforward but extremely tedious if
carried out by hand. For this reason it is a good exercise for
showing the students the advantage and power of computer algebra.
Such software even offer a command for obtaining the inverted
series (\ref{eq:a(v0)_series_exp}) in one step. For sufficiently
small $v_{0}$ this expansion is slightly better than the one
derived in our earlier paper, namely Eq.~(\ref{eq:W0h_weak}), as
follows from comparing the leading terms of the variational and
exact $v_{0}$-series: $W_{0h}\approx -v_{0}^{2}$, $W_{0e}\approx
-1.26v_{0}^{2}$ and $\epsilon _{0}\approx -1.57v_{0}^{2} $.
However, the cubic term that appears in the exact expansion (\ref
{eq:e_weak}) is also missing in the variational treatment based on
the exponential function (\ref{eq:varphi_0e}). From those results
we conclude that $\epsilon _{0}<W_{0e}<W_{0h}$ for sufficiently
small $v_{0}$. In other words, the trial function with the correct
asymptotic behavior at infinity yields a more accurate variational
energy for sufficiently shallow wells.

\begin{figure}[H]
\begin{center}
\bigskip\bigskip\bigskip \includegraphics[width=9cm]{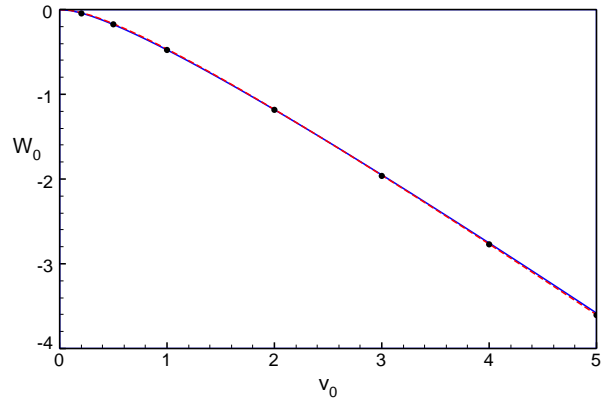}
\end{center}
\caption{(Color online) Variational energies $W_{0e}$ (solid line, blue) and
$W_{0h}$ (dashed line, red) and numerical ones (circles)}
\label{Fig:e0}
\end{figure}

Fig~\ref{Fig:e0} shows $W_{0h}$, $W_{0e}$ and $\epsilon _{0}$ (obtained by
numerical integration\cite{F11b}) for some values of $v_{0}$. The
variational curves are almost indistinguishable in the scale of the Figure.
The numerical results show that they cross at $v_{0}=v_{c}$, where $%
2.4022<v_{c}<2.4023$, so that $W_{0e}<W_{0h}$ if $v_{0}<v_{c}$ and $%
W_{0e}>W_{0h}$ if $v_{0}>v_{c}$. In other words: $W_{0e}$ is more accurate
for shallow wells as argued above and $W_{0h}$ is more accurate for deep
ones. The former inequality is consistent with the previous comparison of
the leading terms of the $v_{0}$-expansions. In the next section we discuss
the deep-well limit.

For the first excited state we propose the simple trial function
\begin{equation}
\varphi _{1e}(a,x)=xe^{-ax},\,x>0  \label{eq:varphi_1e}
\end{equation}
and obtain

\begin{eqnarray}
v_{0} &=&\frac{1}{a\left[ \sqrt{\pi }e^{a^{2}}\left( 4a^{4}+12a^{2}+3\right)
\left[ 1-\mathrm{erf}\left( a\right) \right] -2a\left( 2a^{2}+5\right)
\right] }  \nonumber \\
W_{1e}(a) &=&\frac{a^{2}\left\{ \sqrt{\pi }e^{a^{2}}\left(
4a^{4}+8a^{2}+1\right) \left[ \mathrm{erf}\left( a\right) -1\right]
+2a\left( 2a^{2}+3\right) \right\} }{2\left\{ \sqrt{\pi }e^{a^{2}}\left(
4a^{4}+12a^{2}+3\right) \left[ \mathrm{erf}\left( a\right) -1\right]
+2a\left( 2a^{2}+5\right) \right\} }  \label{eq:W1e(a)}
\end{eqnarray}
The variational energy vanishes at $a_{c}\approx 0.550$ that leads
to the approximate critical parameter $v(a_{c})\approx 1.56$ with
an error of $17\%$, quite smaller than that for the harmonic
function (\ref{eq:varphi_1h}). Once again we appreciate that the
exponential function leads to more accurate results close to
threshold (shallow well).

\begin{figure}[H]
\begin{center}
\bigskip\bigskip\bigskip \includegraphics[width=9cm]{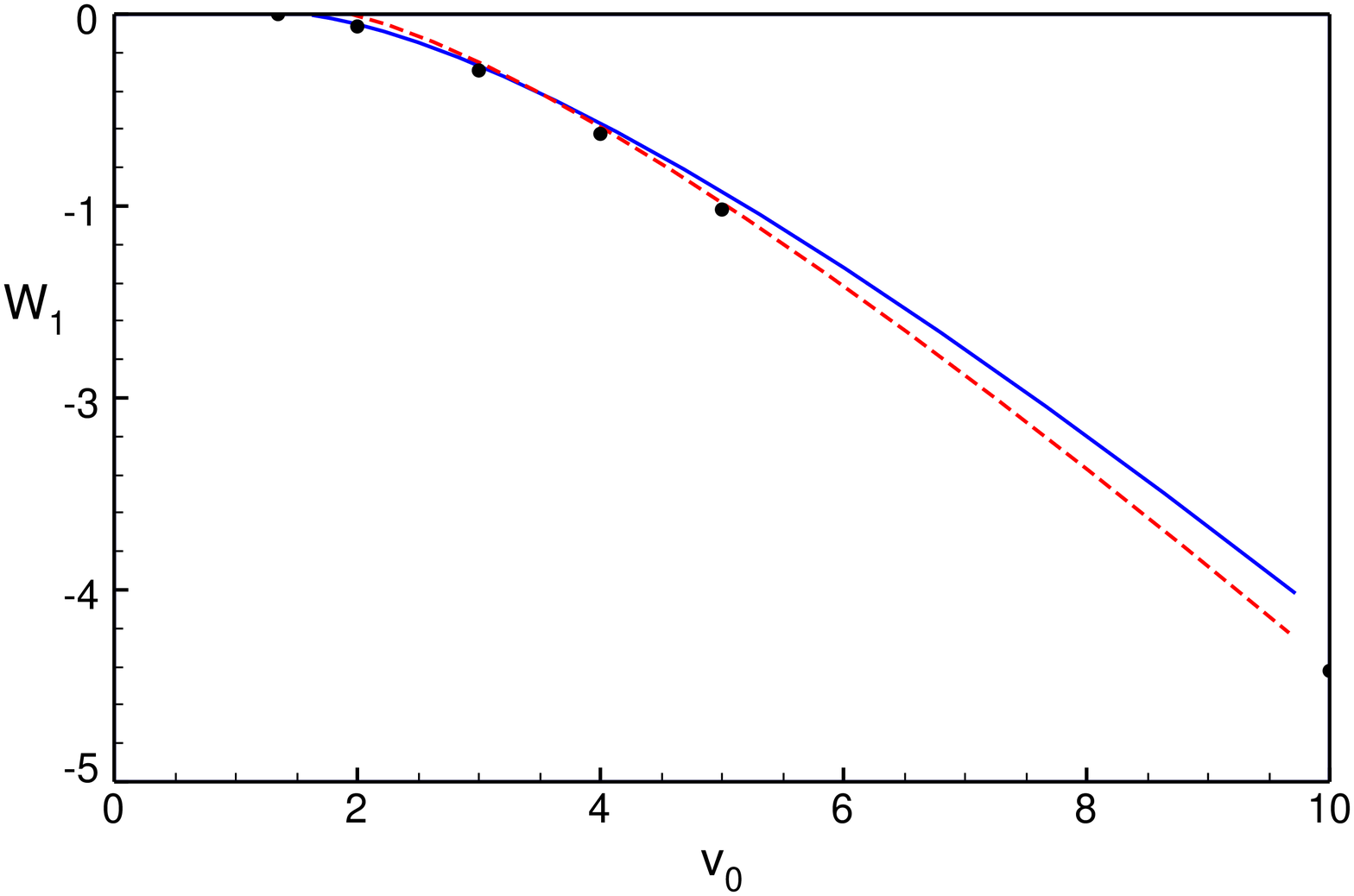}
\end{center}
\caption{(Color online) Variational energies $W_{1e}$ (solid line, blue) and
$W_{1h}$ (dashed line, red) and numerical ones (circles)}
\label{Fig:e1}
\end{figure}

Fig.~\ref{Fig:e1} shows $W_{1h}$, $W_{1e}$ and $\epsilon _{1}$
(obtained by numerical integration\cite{F11b}) for a range of
$v_{0}$ values. In this case the variational curves are not so
close each other and the crossing takes place at
$3.5154<v_{c}<3.51541$. We realize that the exponential function
is preferable for the description of shallow wells as suggested by
the errors in the variational critical well strengths. On the
other hand, the harmonic function leads to more accurate results
for deep wells

\section{The deep-well case}

\label{sec:deep_well}

The deep-well expansion for a general well can be easily derived by means of
perturbation theory.\cite{F00} It follows from those results that the first
terms of the expansion for the Gaussian well are
\begin{equation}
\epsilon _{n}=-v_{0}+\left( n+\frac{1}{2}\right) \sqrt{2v_{0}}-\frac{3}{16}%
\left( 1+2n+2n^{2}\right) +\mathit{O}\left( v_{0}^{-1/2}\right)
\label{eq:exact_deep_exp}
\end{equation}
where $n=0,1,\ldots $ is the harmonic-oscillator quantum number.

In order to obtain the deep-well expansion for $W_{0h}$ first note
that the expression for $v_{0}$ in equation (\ref{eq:W0h(a)})
shows that $a\rightarrow \infty $ as $v_{0}\rightarrow \infty $
and that the leading term is $a\approx \frac{\sqrt{2v_{0}}}{2}$.
Therefore, we substitute
$a=\frac{\sqrt{2v_{0}}}{2}+a_{1}+\frac{a_{2}}{\sqrt{v_{0}}}$ into
the expression for $v_{0}$ and set the unknown coefficients
$a_{1}$ and $a_{2}$ in order to remove the leading powers of
$v_{0}$. In this way we obtain

\begin{equation}
a=\frac{\sqrt{2v_{0}}}{2}-\frac{3}{8}+\frac{3\sqrt{2}}{128\sqrt{v_{0}}}+%
\mathit{O}\left( v_{0}^{-3/2}\right)
\end{equation}
Finally, we substitute this equation into the expression for $W_{0h}(a)$,
expand and keep the leading terms; the result is

\begin{equation}
W_{0h}=-v_{0}+\frac{\sqrt{2v_{0}}}{2}-\frac{3}{16}+\mathit{O}\left(
v_{0}^{-1/2}\right)  \label{eq:W_0h_deep_exp}
\end{equation}
Note that this expression agrees with (\ref{eq:exact_deep_exp}) for $n=0$.

Proceeding exactly in the same way with equation (\ref{eq:W1h(a)}) we obtain

\begin{equation}
a=\frac{\sqrt{2v_{0}}}{2}-\frac{5}{8}-\frac{5\sqrt{2}}{128\sqrt{v_{0}}}+%
\mathit{O}\left( v_{0}^{-3/2}\right)
\end{equation}
and
\begin{equation}
W_{1h}=-v_{0}+\frac{3\sqrt{2v_{0}}}{2}-\frac{15}{16}+\mathit{O}\left(
v_{0}^{-1/2}\right)  \label{eq:W_1h_deep_exp}
\end{equation}
that agrees with (\ref{eq:exact_deep_exp}) when $n=1$. It is clear
that the harmonic variational functions are suitable for the
description of deep wells which is in agreement with the results
in figures \ref {Fig:e0} and \ref{Fig:e1}.

\section{Conclusions}

\label{sec:conclusions}

We have calculated the first two energy levels of the Gaussian
well by means of two types of variational functions. The harmonic
trial functions lead to simpler expressions and their results are
reasonable for all values of the well depth. The approximate
energies exhibit a slightly accurate expansion for shallow wells
and a more satisfactory one for deep wells. In the latter case
they provide the first three dominant terms of perturbation
theory.

On the other hand, the expressions for the energy obtained from
exponential functions are rather complicated and their results are
less accurate for deep wells. However, they are useful if one is
interested in the behavior of the system near threshold (shallow
wells).

Present results suggest that the appropriate behavior of the
wavefunction at infinity does not always guarantee the greatest
accuracy. In the case of a deep well the wavefunction is strongly
localized and compressed about $x=0$ and the description of the
neighborhood of the origin is more important than what happens at
larger $|x|$.

A most intriguing result is that both variational approaches fail
to yield the cubic term in the shallow-well expansion. We have
carried out a similar calculation for the particle in the square
box with finite walls and found that the expansion of the
variational energy given by the trial function
(\ref{eq:varphi_0e}) does not exhibit the cubic term that appears
in the exact series. In this case one can obtain the exact
$v_{0}$-series for $\epsilon _{0}$ from the transcendental
equation that determines the energy levels of that model. We do
not have a satisfactory explanation for this failure of the
variational method.

\end{document}